\documentclass[aps,twocolumn,showpacs,superscriptaddress]{revtex4}
\usepackage[dvips]{graphicx}
\usepackage{amsmath} 
\usepackage{amssymb}
\usepackage{dcolumn}
\usepackage{pifont}

\begin{document}
\title{Crucial role of sidewalls in velocity distributions in quasi-2D granular gases}
\author{J. S. van Zon}
\affiliation{Center for Nonlinear Dynamics and Department of Physics, University of Texas, Austin, Texas, 78712}
\affiliation{Division of Physics and Astronomy, Vrije Universiteit, 1081 HV Amsterdam, The Netherlands}
\author{J. Kreft}
\email[Email address: ]{kreft@chaos.utexas.edu}
\affiliation{Center for Nonlinear Dynamics and Department of Physics, University of Texas, Austin, Texas, 78712}
\author{Daniel I. Goldman}
\affiliation{Center for Nonlinear Dynamics and Department of Physics, University of Texas, Austin, Texas, 78712}
\author{D. Miracle}
\affiliation{Center for Nonlinear Dynamics and Department of Physics, University of Texas, Austin, Texas, 78712}
\author{J. B. Swift}
\email [Email address: ] {swift@chaos.utexas.edu}
\affiliation{Center for Nonlinear Dynamics and Department of Physics, University of Texas, Austin, Texas, 78712}
\author{Harry L. Swinney}
\affiliation{Center for Nonlinear Dynamics and Department of Physics, University of Texas, Austin, Texas, 78712}

\date{\today}
\begin{abstract}

Our experiments and three-dimensional molecular dynamics simulations
of particles confined to a vertical monolayer by closely spaced
frictional walls (sidewalls) yield velocity distributions with
non-Gaussian tails and a peak near zero velocity. Simulations with
frictionless sidewalls are not peaked. Thus interactions between
particles and their container are an important determinant of the
shape of the distribution and should be considered when evaluating
experiments on a constrained monolayer of particles.

\end{abstract}

\pacs{81.05.Rm, 45.70.-n, 05.20.Dd, 05.70.Ln}

\maketitle

Granular materials can mimic the behavior of different states of
matter, including a gas\cite{kadanoff,williams,biot,goldhirsch}. Since
collisions with grains are inelastic, the gaseous steady state can
only be maintained by external forcing. Despite much recent work, the
form of the velocity distribution for a driven granular gas remains an
open question --- velocity distribution functions found in experiment
\cite{mustard, nature, urbach, urbach2, urbach3, losert, kudrolli,
kudrolli3, wolpert, aranson, rouyer}, simulation \cite{aranson, moon,
pagnani, barrat, barrat2, brey}, and theory
\cite{ernst,BNK,antal,puglisi} differ significantly.

The velocity distribution function for elastic particles in
equilibrium is Gaussian. Distributions obtained for inelastic granular
gases are typically not Gaussian and are often fit to a function of
the form
 \begin{equation}
  P_\alpha(v)=a\exp(-B {|v/\sigma|}^\alpha)
 \label{eqno1}
 \end{equation}
 where $a,B$ and $\alpha$ are fitting parameters, and $\sigma=\langle
 v^2\rangle^{1/2}$ \cite{mustard, losert, aranson, rouyer, barrat2,
 brey, moon}. Several experiments and simulations with different
 geometries and forcing mechanisms have found $\alpha \approx 1.5$
 \cite{losert,aranson,rouyer,barrat2} although Gaussian distributions
 ($\alpha=2$) have also been observed \cite{nature,
 urbach3}. Simulations have revealed distributions that are not
 described by a single function but instead display a crossover from
 $\alpha=2$ at low velocity to $\alpha<2$ at high velocities
 \cite{brey,moon}. The value $\alpha=3/2$ has been obtained for the
 large velocity limit for the special case of a gas of inelastic
 frictionless particles with homogeneous stochastic forcing and no
 gravity~\cite{ernst}.

Many experiments have been conducted on monolayers of particles
\cite{urbach,urbach2,urbach3,losert,kudrolli,kudrolli3,wolpert,rouyer}
because limiting the motion in one dimension allows the use of a video
camera to record the entire velocity field.  Since velocities in the
suppressed direction can never be fully eliminated, these systems are
quasi-2-dimensional (2D), not strictly 2D. In such confined geometries,
particles can make as many or more collisions with the wall as with
other grains during one driving cycle.  Collisions with walls may then
influence the shape of the velocity distribution function.  We find
that the confining sidewalls play a major role in determining the
velocity distribution function, which we obtain from experiments and
simulations on a vertically oscillating monolayer of spheres
whose motion is suppressed in one horizontal direction.

\begin{figure} [hb]\centering

\includegraphics[width=8cm]{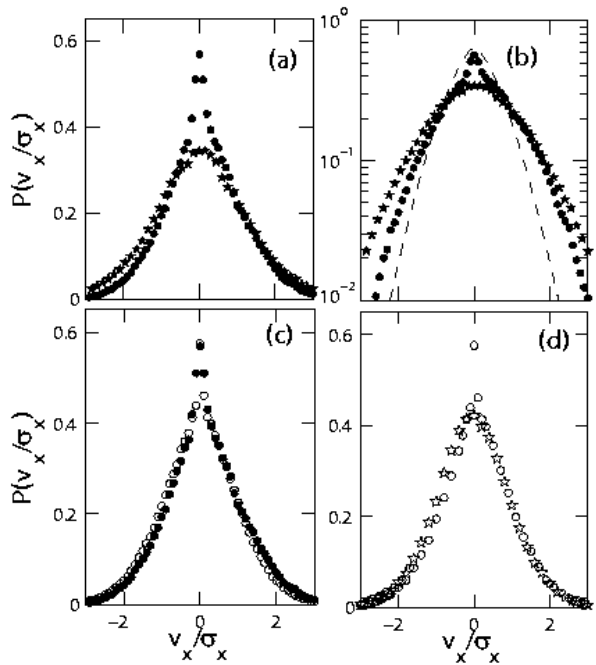} \caption{(a) Linear and (b)
  logarithmic plots of the velocity distribution $P(v_x/\sigma_x)$
  measured in the steady state region for a system with $N=130$,
  $f=50$Hz and $\Gamma=20$. Distributions are
  shown for clean particles ($\bullet$) and particles with a small
  amount of added graphite ($\star$). Also shown is the non-Gaussian
  result (eq. 1, dashed line) from the experiment in ~\cite{rouyer},
  with $\alpha=1.51$ and $B=0.8$. (c) Comparison between experiment
  ($\bullet$) with particles and simulation
  ($\circ$) with $\mu_w=0.075$. (d) Comparison of simulations with
  ball-ball friction $\mu_b=0.5$ and with ball-wall friction
  $\mu_w=0.075$ ($\circ$) and $\mu_w=0$ (\ding{73}). The experimental
  distributions are not precisely symmetric about $v_x=0$ due to the
  container tilting slightly when shaking. To match the asymmetry in
  the experiment, gravity in the simulation was tilted 1.9 degrees
  with respect to the normal to the top of the container. This does
  not affect the functional form of the distributions when compared to
  simulations without the tilt.}\label{fig:figure1}
\end{figure}

\emph{Experiment.} Our experimental setup, which is similar to that in
\cite{rouyer}, used $N=130$ stainless steel balls of diameter $d=1.6$
mm, contained between vertical sidewalls (Plexiglas plates) with a
separation of $1.1d$. The container had an interior horizontal
dimension of $48d$ and vertical dimension of $32d$. It oscillated with
a frequency $f=50$ Hz, and the peak non-dimensional acceleration was
$\Gamma=4\pi^2f^2A/g=20$, which corresponds to an amplitude
$A=1.25d$. The container was evacuated ($8$ Pa) to avoid hydrodynamic
interactions. Each run used new balls that were cleaned in ethanol and
sonicated. In our experiment particles gained energy only through
collisions with the bottom of the container (in earlier measurements
at $\Gamma=50$~\cite{rouyer}, particles collided with the top as well
as the bottom of the container, which was not evacuated). Particle
motions were recorded by a digital camera (Phantom v4, Vision
Research) at a rate of $1000$ frames per second. Particle
displacements were resolved with an accuracy of $~0.004$ mm
($0.0025d$). Statistical properties were obtained by averaging over
$7650$ drive cycles and 20 different phases in the cycle.

\emph{Simulation.} An event-driven algorithm described in \cite{bizon}
was used for the simulation, which was conducted for the same
$\Gamma$, $f$, and sidewall separation as the experiment. The
parameters characterizing ball-ball interactions were the minimum
coefficient of restitution $e=0.7$, the coefficient of sliding
friction $\mu_b=0.5$, and the rotational coefficient of restitution
$\beta=0.35$. The coefficient of restitution varies with relative
normal velocity ($v_n$) as described in \cite{bizon}: the
coefficient of restitution is the maximum of $e$ and
$1-(1-e)(v_n/\sqrt{gd})^{3/4}$. The TC model of Luding and McNamara
\cite{luding} was also used to prevent inelastic collapse by setting
the coefficient of restitution to unity if a particle was involved in
another collision within $3.7 \times 10^{-4}$ seconds of the previous
one. For interactions between balls and the container (both the
sidewalls and bottom), we used the same values for $e$ and $\beta$,
but we varied the coefficient of sliding friction with the wall from
$\mu_w=0$ (no sidewall or bottom friction) to $\mu_w=1$. To reproduce
the experiment, $N=130$ particles were simulated in a box of height
$200d$, width $48d$, and plate separation $1.1d$. The entire box was
oscillated vertically so the particles collided with moving sidewalls,
in addition to the bottom, as in the experiment.

\emph{Steady-state distributions.} Collisions of particles with the bottom plate inject energy mainly into vertical motion. Energy is transferred into the horizontal direction directly through particle-particle collisions and through collisions of rotating particles with the bottom. Close to the bottom plate the areal density $\rho$ and the probability distributions for the horizontal and vertical components of velocity ($v_x$ and $v_z$) vary considerably during each oscillation of the plate.  However, far above the plate the density and velocity distributions become time independent, as has been shown by Moon \emph{et al.} ~\cite{moon}. Here we examine distribution functions for $11d<z<12d$, which is in the steady state region -- the density and horizontal velocity distribution functions change by less than $5\%$ during each cycle.

Our measured and simulated distributions are shown in
Fig. \ref{fig:figure1}. For clean particles in
the experiment and non-zero wall friction in the simulation, the
velocity distributions have an unusual characteristic: a sharply
peaked maximum, a feature that has been observed before \cite{urbach,
kudrolli} but has not been fully explored. (For $z<11d$, the shape of
the distribution changes slightly with height in the box and phase of
the driving cycle, but the sharp peak is always present.) We find that
the peak disappears when we add approximately $0.0002$ kg of graphite
powder (a lubricant) to the $0.1$ kg of steel
spheres~\cite{goldman}. The distributions observed with and without
graphite both differ from those in \cite{rouyer}
(cf. Fig.~\ref{fig:figure1}(b)).

Experiment and simulation are compared in
Fig. \ref{fig:figure1}(c). For $\mu_w=0.075$, the simulation results
agree well with the experiment. The peak of the velocity distribution
in the simulation decreases as $\mu_w$ is decreased, and the peak
disappears completely for $\mu_w=0$, as Fig. \ref{fig:figure1}(d)
shows.
 
\begin{figure} [hb]\centering
\includegraphics[width=8cm]{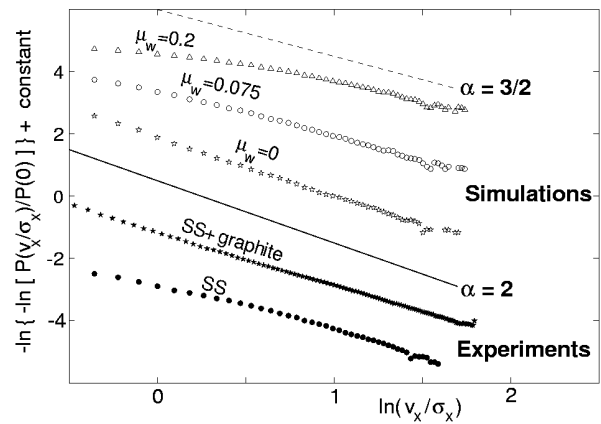} \caption{Double log plot of the
velocity distribution functions.  $P_\alpha(v)$ has slope $-\alpha$;
to guide the eye, slope $\alpha=3/2$ is shown by a dashed line and
$\alpha=2$ (a Gaussian) by a solid line. Experimental results are
shown for clean stainless steel (SS) particles ($\bullet$) and SS
particles with graphite added ($\star$). Simulation results are shown
for three different values of ball-wall friction $\mu_w$ with the
ball-ball friction held fixed, $\mu_b=0.5$. The data sets have been offset
for clarity.}
\label{fig:figure2}
\end{figure}

The distributions obtained from experiments on stainless steel
particles with graphite and simulations with $\mu_w=0$ are described by
a straight line on graphs like those in
Fig. 2.  The slope of such a graph yields the magnitude of the
exponent $\alpha$ in $P_\alpha(v)$.  In simulations without sidewall
friction, $\mu_w=0$, the exponent obtained is $1.8$.  An exponent of
$1.7$ is found for the velocity distribution of stainless steel
particles with graphite.  The peaked distributions are not described
by a single value of $\alpha$, but we can compare estimates of
a local value of $\alpha$ in the range $1.0\lesssim ln(v_x/\sigma_x)\lesssim
1.6$:  we obtain 1.8 for clean stainless steel particles, while in the
simulation, $\alpha$ increases from 1.3 with sidewall friction 
$\mu_w=0.2$ to $\alpha=1.8$ with $\mu_w=0.1$. 

\begin{figure} [hb]\centering
\includegraphics[width=8cm]{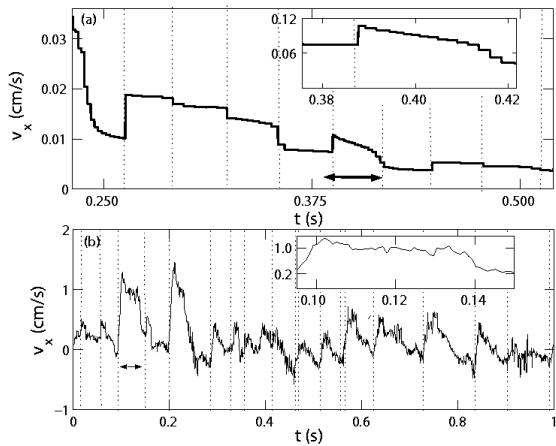} \caption {Horizontal velocity of
a \emph{single} ball on a vertically oscillating plate in (a)
simulation ($\mu_w=0.4$ and $\mu_b=0.5$) and (b) experiment. The
larger, less frequent jumps are the result of collisions with the
bottom plate; these collisions are indicated by the dotted vertical
lines. The more frequent smaller changes are the result of collisions
with the sidewalls; these changes are clearer in the enlarged scale of
the insets. The regions depicted in the insets are indicated by
arrows.}
\label{fig:figure3}
\end{figure}

\emph{Single Particle Dynamics.} We have shown that interactions with
sidewalls strongly affect the functional form of the velocity
distribution. This result is supported by our observations in
simulations that in the steady state region, a ball collides with the
wall typically three times as often as it collides with another ball.
To isolate the effects of ball-wall collisions, we have conducted
experiments and MD simulations on a single particle in an oscillating
container (Fig. \ref{fig:figure3}). Because there are no collisions
with other particles, the particle's motion is determined only by
collisions with the bottom plate and the sidewalls. Figure
\ref{fig:figure3}(a) shows the time evolution of the horizontal
velocity $v_x$ for a particle in a simulation with $\mu_w=0.4$ and
$e=0.7$.  Each time a particle bounces on the bottom plate, some of
the angular momentum of the particle can be transferred into linear
momentum in the horizontal direction. These collisions would produce
the only changes in $v_x$ if there were no interaction with the
sidewalls, but Fig. \ref{fig:figure3}(a) reveals more frequent smaller
changes, which correspond to collisions with the sidewall. The
staircase-like decrease in velocity (see inset) corresponds to a
particle's rattling between the sidewalls, losing energy at every
collision. Thus the effect of the sidewalls is to damp the horizontal
velocity. The ultimate fate of a single particle, regardless of its
initial $v_x$, is to bounce vertically on the bottom plate with
$v_x=0$.

The horizontal velocity $v_x$ measured for a single particle in the
experiment is shown in Fig. \ref{fig:figure3}(b). Collisions with the
bottom plate, determined to be when the vertical component of velocity
$v_z$ changes sign, are indicated by the dotted vertical lines. If
there were no influence of the sidewalls, the horizontal velocity
$v_x$ would remain constant between these lines. The behavior of the
particle between collisions with the bottom plate is more complicated
than in the simulation, but it is still clear that the horizontal
velocity is damped by collisions with the walls. The damping of the
horizontal motion of a single particle illustrated by
Fig. \ref{fig:figure3} explains why the velocity distribution for a
gas of particles has a peak at $v_x=0$ (Fig. \ref{fig:figure1}).  The
over-populated high energy tails arise because for a distribution with
a given variance, the increase in the central peak must be balanced by
an increase for $v>\sigma$.

\begin{figure} [hb]\centering
\includegraphics[width=8cm]{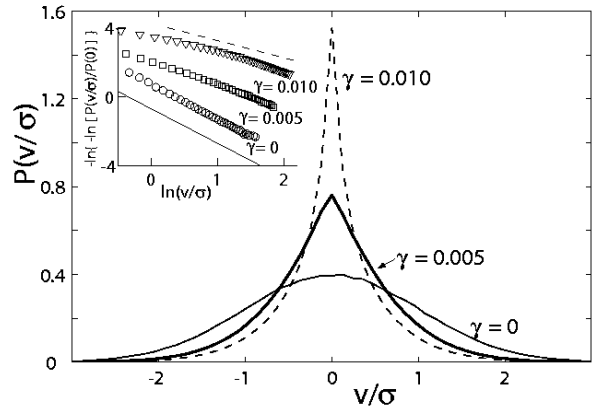} \caption{Velocity distribution
  $P(v/\sigma)$ for a model of a damped driven particle. The
  distributions are shown for increasing damping $\gamma$: 0, 0.005,
  and 0.010. The inset compares the tails of the distributions for the
  model with a Gaussian distribution (solid line, $\alpha=2$) and a
  distribution with $\alpha=1.5$ (dashed line). The data sets are
  offset for clarity.} \label{fig:model}
\end{figure}

\emph{Single Particle Model.} Features of the velocity
distributions obtained from experiment and simulation are well
described by a discrete map model with a damped driven single
particle.  The particle's velocity is initially drawn from a Gaussian
distribution of variance unity. The velocity at iteration {\it n+1} is
given by $v_{n+1}=v_n e^{-\gamma}$. For one percent of the iterations,
randomly selected, we replace the velocity {\it $v_{n+1}$} with a
velocity drawn from a Gaussian distribution with variance unity.  The
velocity probability distribution is constructed from $n=10^9$
iterations. The exponential decay of the particle velocity between
iteration steps corresponds to the numerous sidewall collisions that occur
between excitations by the plate, and the random replacements of the
particle's velocity mimic plate collisions that transfer horizontal
momentum to the particle.

This model captures the qualitative behavior of the velocity
distributions found in both experiment and simulations, as
Fig. $\ref{fig:model}$ illustrates. For finite damping, $\gamma>0$,
the distribution is strongly peaked at $v=0$, while in the absence of
damping, $\gamma=0$, the distribution is Gaussian. Further, damping
affects the tails of the distribution: as damping is decreased to
zero, double logarithmic plots of the distribution become less curved
and the slope increases from 1.3 to 2, just as in the MD simulation
(Fig. \ref{fig:figure2}).


The single particle model is similar to a model by Puglisi \emph{et
al.} \cite{puglisi} that includes damping of the particle velocities.
Increasing the damping in their model also led to non-Gaussian
velocity distributions, but a strong peak around $v=0$ was not
reported.  This peak might be absent in their model because particles
were driven not by discrete heating events but by continuous white
noise, which for strong damping led to Gaussian behavior around $v=0$ in their model.

\emph{Conclusions.}  The kinetic theory of granular gases is often
studied in experiments on confined monolayers of grains because the
behavior of all grains for all times can be recorded. However, we have
found that the ball-wall friction associated with the confinement
should be included in interpreting experiments on monolayers
in quasi-2D geometries, including vertical \cite{rouyer}, inclined
\cite{kudrolli}, and horizontal layers \cite {urbach, nature}. Indeed,
in an experiment with the last geometry the velocity distribution was
peaked for a smooth plate~\cite{urbach}, but the peak disappeared when
the smooth plate was replaced with a rough plate, which drove
horizontal as well as vertical motion~\cite{urbach3}. Similarly, a
recent experiment with a layer of light particles on top of a layer of
heavy particles yielded a non-Gaussian distribution for the heavier
particles, but Gaussian statistics were found for the lighter
particles \cite{nature}. The interactions between the particles and
the container in these quasi-2D systems may have been principal
determinants of the shape of velocity distributions and therefore should
be taken into consideration.

We thank Fred MacKintosh, Sung Joon Moon, and Erin Rericha for helpful
discussions.  This work was supported by the Engineering Research
Program of the Office of Basic Energy Sciences of the U.S. Department
of Energy (Grant No. DE-FG03-93ER14312), The Texas Advanced Research
Program (Grant No. ARP-055-2001), and the Office of Naval Research
Quantum Optics Initiative (Grant N00014-03-1-0639).

\end{document}